\documentclass[a4paper,11pt]{article}
\pdfoutput=1
\usepackage{amssymb,amsmath,amsfonts,makeidx,placeins,pbox,multirow}
\usepackage{graphicx,rotate,subcaption,color,slashed,cite,caption,epstopdf,verbatim}
\usepackage{longtable,tabu}
\usepackage{array}
\usepackage{axodraw4j}
\usepackage{pstricks}
\usepackage{color}
\usepackage{amsmath}
\usepackage{mathtools}
\usepackage{amssymb}
\usepackage[normalem]{ulem}
\newcolumntype{P}[1]{>{\centering\arraybackslash}p{#1}}
\usepackage[colorlinks=true,
            linkcolor=magenta,
            urlcolor=blue,
            citecolor=blue]{hyperref}

\evensidemargin=\oddsidemargin
\def\bar {\overline}

\def\bea {\begin{eqnarray}}
\def\eea {\end{eqnarray}}

\def\beq{\begin{equation}}
\def\eeq{\end{equation}}
\def\barr{\begin{array}}
\def\earr{\end{array}}

\def\gev{\ensuremath{\mathrm{Ge\kern -0.1em V}}}

\begin{document}
\begin{center}
{\Large \bf A Brief Review on Jet Substructure in Connection With\\ \vspace{10pt} 
Collider Phenomenology}
\\
\vspace*{0.5cm} {\sf Nilanjana Kumar} \footnote{nilanjana.kumar@gmail.com} \\
\vspace{10pt}
{\small Centre for Cosmology and Science Popularization, SGT University, Gurugram, Delhi-NCR, India} \\
\normalsize
\end{center}
\bigskip
\begin{abstract}
It is a challenge for the theoretical particle physicists to perform 
the phenomenology of the Beyond Standard Model (BSM) theories using advanced 
simulations which can mimic the experimental environment at the colliders as closely as possible. 
In collider phenomenology {\it jet substructure} is a concept that is used 
frequently to analyse the properties of the jets, characterised 
as a cluster of hadrons, which is often the end result of particle collisions at 
the colliders, such as Large Hadron Collider (LHC).
A vast literature on jet substructure exists both from the 
theory and experimental point of view. But, even with the knowledge 
of Quantum Chromodynamics (QCD), it is hard to cope up with the vastness 
of the applicability of the subject for a new researcher in this field.
In this review, an attempt has been made to bridge the gap between the concept
of jet substructure and its application in the collider phenomenology. However, 
for detailed understanding, one should look at the references.
\end{abstract}
\tableofcontents
\section{Introduction}
Even though there are experimental hints for physics 
beyond Standard Model (BSM), no direct evidence of the new physics has been observed. 
There exists a hierarchy between the scale of new physics and the
electroweak scale. Also, the new physics may contain traces of new particles at TeV scale or 
beyond. If these new particles exist, they are produced at the high energy collisions 
along with SM particles. Interactions of the new particles 
with the Standard Model (SM) particles such as 
bosons ($W$, $Z$, $H$) and/or fermions are governed by the underlying BSM theories. Hence the BSM 
particles decay to the SM particle and 
when these SM particles decay, they produce hadrons, leptons or 
radiation in the final state.  
{\it Jets} are termed as colimated bunch of hadrons resulting from
quarks and gluons produced at high energy. {\it Jet substructure} is an array 
of tools to extract information from the radiation pattern inside the jets. 

It is very interesting to see that jet substructure covers 
information about a vast span of scale: it is sensitive to probe the 
new physics at $\Lambda_{new}$ but it obeys the principals of 
QCD at scale  $\Lambda_{QCD}$. The electroweak scale $\Lambda_{EW}$ on the other hand 
lies between these two. A sufficient condition
to ensure calculability within the perturbation theory of QCD is Infrared and
Collinear (IRC) safety with many others, which has therefore played a central role as an organizing
principle for jet observables. For a brief history and theoretical development in 
jet substructure one might look at Ref:~\cite{Marzani:2019hun,Abdesselam:2010pt,Altheimer:2012mn,Adams:2015hiv}.

Current limits on the new BSM particles is getting larger than about 1 TeV, 
as inferred from the data gather by the Large Hadron Collider (LHC) experiment.
If the BSM particles are very heavy ($> 1$TeV), its decay produce are highly colimated.
For example the bosons or the top quark coming from the 
decay of a new particle will be highly boosted and when 
they further decays to leptons or jets with high transverse momentum ($p_T$). The 
decay products are highly colimated and hard to distinguish. In these scenarios, 
jet substructure is used to identify boosted hadronically decaying electroweak
bosons and top quarks instead, which in turn improves the sensitivity for the new
physics searches. Researchers are developing new theoretical ideas
and reconstruction techniques to probe the jet substructure at high 
energy frontier of LHC.

In this review, I briefly summarize the QCD aspects of jet substructure in section 2.
How a jet is reconstructed is described briefly in section 3.
In section 4, I discuss different aspects of the jet substructure variables, techniques 
with example of a BSM scenario. After which I conclude by discussing some more modern aspects of 
jet substructure.
\section{Jet Substructure: QCD Point of View}

The hadrons in the SM is be described by the Quark Model ~\cite{Gell-Mann:1964ewy} developed by 
Murray Gell-Mann, Kazuhiko Nishijima and George Zweig. 
If the nucleus of a particle is bombarded with electrons, such as in 
Deep Inelastic Scattering Experiments (DIS)~\cite{RevModPhys.67.157}, the proton shows 
the behaviour of a point like objects. The constituents of 
the proton is described the Parton Model~\cite{Feynman:1969ej,Bjorken:1969ja}, developed by Richard Feynman. 
Parton model assumes that at high energies, the hadrons can be seen 
as made up of other free constituents (quarks and gluons), 
which carries a fraction of its momentum. 
Quantum Chromo Dynamics (QCD)~\cite{Gross:2005kv,Wilczek:2005az} is the theory of strong 
interactions describing the interactions of quarks and gluons.
A detailed review of QCD in the context of jet substructure can be found in Ref:\cite{Marzani:2019hun,ellis_stirling_webber_1996}
and the references within. 
I will discuss only some important aspects of QCD in the 
following, which is the necessary building block of jet substructure observables.
\subsection{QCD and Asymptotic Freedom}
The perturbative calculation of QCD requires the strong coupling constant $\alpha_s$ to be small, 
in order to allow the perturbative techniques to produce efficiently at short distances, i.e, 
at high energies. It requires some sort of renormalization in order to remove the Ultraviolet Divergences (UV)
, which in turn introduces a cut-off scale ($\mu$). The renormalization of the gauge coupling $\alpha_s$ 
depends on $\mu$ and the bare coupling  $\alpha_s (\mu_0)$. The solution of 
the Renormalization Group Equation (RGE) \cite{collins_1984} at one loop can be written as  ,
\begin{equation}
\alpha_s (\mu) = \frac{\alpha_s(\mu_0)}{\big(1 - \frac{b_0 \alpha_s (\mu_0)}{2\pi} \log(\frac {\mu}{\mu_0})\big)}\\
\end{equation}
where, $b_0 = -11+ (2/3) n_f$. Now, introducing the QCD scale $\Lambda_{QCD}=\mu_0 \exp{\frac{2\pi}{b_0 \alpha_s(\mu_0)}}$,
$\alpha_s$ takes the following form.
\begin{equation}
\alpha_s (\mu) = \frac{2\pi}{b_0 \log(\frac{\mu}{\Lambda_{QCD}})}
\end{equation}
It implies that at $\mu=\Lambda_{QCD}$, $\alpha_s$ blows up, even if Higher order corrections \cite{Czakon:2004bu}are included. 
Hence, at smaller energies, the predictions of QCD is not totally reliable. On the other hand, at a larger energy, 
$\alpha_s$ gets smaller and smaller, implying that the results of the perturbative theory will be more accurate 
at higher energies. In other words, this means that quarks are almost ``free'' particle, at high energy, 
and this feature is termed as ``Asymptotic Freedom". For a more detailed discussion we refer to \cite{RevModPhys.77.837}.
\subsection{Parton Model and Parton Density Functions(PDF)}
A hadron is a bound state of partons, which consists of quarks, anti-quarks and gluons. 
The confinement scale is $1/\Lambda_{QCD}\sim 10^{-13}$ cm.
The scattering of a parton consists of two processes. (1) {\it Hard Scattering}: 
When a point-like parton is scattered, for example, $q \bar{q} \rightarrow q \bar{q}$. 
(2) {\it Soft Scattering}: Radiation ($\gamma$) and exchange of virtual gluon ($g$) with low energy.
In order to calculate the cross section of the hard scattering process (in pp collision), it is required to know what 
is the fraction of energy the initial state parton carries. This is manifested in Parton Density Functions.
The detailed descriptions of the quark anti-quark and gluon PDF's can be found in \cite{Ethier:2020way}. 
The PDF's are universal and do not depend on a particular process, and they are determined 
by fitting data from several experiments. Hence, It is very essential to match these PDF's 
with the experimental data. The current status 
of various PDF's is given in Ref:\cite{Rojo:2015acz}. 

If a parton $k$, carries $x$ fraction of a hadron, then, the probability of finding a parton of 
type $k$ with 4 momentum between $xp^{\mu}$ and  $(x+dx)p^{\mu}$ inside a hadron with 
four momentum  $p^{\mu}$ is given as, $f^k(x) dx$. This is the parton density function for $k$.
Hence, PDF's are function of the momentum fraction ($x$) and also another parameter, {\it Factorization Scale ($Q$)}.
This scale can be thought of as a boundary where short distance physics ends and long distance physics 
takes over. In some cases, the factorization scale is the same as {\it Renormalization Scale ($\mu$)}, but that is 
not the case always. The dependence can be described in terms of DGLAP equations \cite{Martin:2008cn} in the perturbation 
theory. Here in analogy with the running of the gauge coupling ($\alpha_s$), the factorization scale is varied and 
the RGE's of the PDF's are obtained.
The cross section for a collision would be technically be independent of these two scales in a hypothetical 
scenario, when all order of corrections are included \footnote {This is a very important 
cross check (often forgotten in many literature) in phenomenology while generating any process, say in Madgraph}. 
But in reality, the fixed order calculation of cross section 
shows dependence on both the scales. The differences between the dependencies minimizes, 
at higher order calculation of perturbative QCD.
\subsection{Perturbative Calculation in QCD and Some Concerns}
Perturbative calculation in QCD gives theoretical precision by calculating the 
higher order Feynman diagrams in order of $\alpha_s$. The loop diagrams induce two 
types of singularities:\\

\paragraph {(1) Ultraviolet singularities (UV):} This type of divergences occur when the loop momentum goes 
to infinity. It can be regularised as QCD is a renormalizable theory and it 
leads to renormalized wave function and running couplings~\cite{Lovett-Turner:1994rmm}.

\paragraph{(2)Infrared and collinear singularities (IRC):\cite {PhysRevLett.39.1436}} 
This type of singularities arise from interactions that happen long time after 
the creation of the particles and that perturbation theory in QCD breaks down in
long-time physics. But the detector is situated a long distance away from
the interaction point. Hence somehow we have to account for the 
long distance physics (IR) in the theory.

This type of divergence occur when 
a parton is split into two collinear partons that is with parallel four momenta ($k_\mu$) 
or a soft parton with $k_\mu\rightarrow 0$ is added with a parton.
Let us consider a particle 1 to split in parton 2 and 3, where 2 gives rise to hard scattering or hadronization.
Now, one gets the relation
\begin{equation}
\frac{1}{k_2^2}=\frac{1}{-2E_1 E_3(1-\cos\theta)}
\end{equation}
Now, either $E_3=0$ or $\theta=0$ give divergence, and that leads to singularity. 
When $E_3=0$, i.e, the particle 3 has a very small energy, 
the singularity is called soft singularity and the emission of 
3 is characterised as {\it soft emission}. 
The limit $\theta=0$ corresponds to ''{\it Collinear singularity}''.
Now, if the phenomenon is characterised by Quantum Electrodynamics (QED), for example, emission of $Z$
to two fermions, these divergences cancel out with the additional process, where an external soft photon is emitted 
($Z\rightarrow \gamma\gamma$). This is explained by Bloch-Nordsieck theorem~\cite{Bloch:1937pw}.
But such cancellation is not possible in QCD, when dealing with hadrons only.

Appearance of infrared divergence indicates that the calculation depends on the long distance 
aspects of QCD. This problem can be resolved by if we restrict ourself to the {\it infrared and collinear safe} 
observables. Another thing which is doable is separating out the non negligible non perturbative effects 
in the calculation of the variables, for example the cross section, where, \\
\begin{equation}
\sigma=\sigma_{pert}+ \sigma_{non pert}.
\end{equation}
In order to do that, {\it Splitting Functions} are useful, which is based on the 
{\it Weizsacker-Willims} spectrum of QCD. 
Moreover, based on that, one can obtain the correct set of evolution equations for the PDF's, 
which are known as Altarelli-Parisi equations~\cite{Borah:2012ey}. 

Another important thing to mention here is that it is well established that fixed order calculation 
of the cross section fails in the particular limit of the phase space, hitting singularities. 
Hence it is good idea to 
use all order calculation, where emission of particles and particles in the loops are considered. But 
again these calculations highly depend on the kinematics of the particular event. Hence, in order 
to get a reliable physical value of say $\sigma$, one needs to match these two approaches. 
This is done by the {\it Matching Schemes}~\cite{Frixione:2002ik,Frixione:2007vw} in the event generators \footnote {More on that from \href{https://portal.nersc.gov/project/alice/pythia8doc/htmldoc/JetMatching.html}{here}.}. 
\begin{equation}
\sigma_{matched}=\sigma_\text{fixed order} + \sigma_\text{resummed} - \sigma_\text{double counting}
\end{equation}

\paragraph{IRC safe Observables:} An observable, such as cross section is ``IRC safe'' means that 
the value of the observable remains unchanged so that proper cancellation of the divergence coming 
from the virtual particle emission or loop diagrams is achieved for the ensemble of soft or collinear 
partons. Even though the singularities cancel, the kinametic dependence of the observables can 
cause an imbalance between the real and virtual contributions leading to large logarithmic terms in any order 
of perturbation theory. As a result, the perturbative expansion of the strong couplings are spoiled. This is
achieved by re-summing the contribution to all orders. One such example where it occurs is the observable 
{\it jetmass} in high $p_T$ region. Here the problem is looked after 
by the jet substructure algorithms called {\it Groomers}~\cite{Krohn:2009th,Ellis:2009me,Larkoski:2014wba}.
Algorithms such as anti-$k_T$~\cite{Cacciari:2008gp} is IRC safe whereas PDF's ~\cite{Soper:1996sn} are not IRC safe.

In many cases the IRC safe observables are also insensitive to arbitrary soft gluon emission or collinear 
parton splitting. Those IRC unsafe observables has also been studied by the help of 
{\it perturbative Sudakov Form Factor}. 
The mechanism regulates the real and virtual IR divergences \cite{Larkoski:2015lea}.
One example of such characteristic is the {\it momentum fraction} 
$z=min(p_{T1},p_{T2})/(p_{T1}+p_{T2})$. 
It can be shown that this quantity in modified mass drop tagger (mMDT)~\cite{Dasgupta:2013ihk} is Sudakov safe 
in some regions.
\section{Jet Reconstruction}
The quarks and gluons are not observed as the final state of a collision. 
They tend to decay to other quarks and gluons and finally produce 
a colimated bunch (radius of the cone is very small) 
of quarks and gluons which we call {\it jet}. Some of them 
combine and form hadrons. 
These final state particles, which are fragments of the main jet are called ({\it subjets}). These subjets 
are combined in such a way that the 
initial jet can be reconstructed. The basic variables to define
the kinematics of a jet is {\it jet radius} ($R$)
which is a function of the {\it rapidity} ($y$) and {\it azimuthal angle} ($\phi$). 

Before getting into the jet, let us define some useful quantities first.
Consider the transverse momentum of the jet $(E, p_x, p_y, p_z)$ to be $p_T$, where,
\begin{equation}
p_T=\sqrt {(p_x^2+p_y^2)}
\end{equation}
The distance between two jets are given by, 
\begin{equation}
\Delta R_{ij}=\sqrt {(\Delta y_{ij}^2+\Delta \phi_{ij}^2)}
\end{equation}
where the rapidity is defined as,
\begin{equation}
y=\frac{1}{2}\log \frac{E+p_z}{E-p_z}
\end{equation}
Another important variable from the experimental point of view is, 
{\it pseudo-rapidity} ($\eta$), defined as,
\begin{equation}
\eta=\frac{1}{2}\log \frac{|\vec p|+p_z}{|\vec p|+p_z}=\log \tan(\theta/2)
\end{equation}
where $\theta$ is the {\it polar angle} in the beam direction.

\paragraph{Jet Algorithms:}
Jets are identified via the {\it jet algorithms}~\cite{Salam:2010nqg} which is broadly divided into 
two classes:
{\it Sequential recombination algorithm} and {\it Cone algorithm}.

In sequential recombination algorithm, two daughter jets which are close by are selected and 
combined to form the mother jet and 
this process goes on until the direct decay product of the collision is traced. 
The examples of this type of popular algorithms are :
\begin{itemize}
\item $k_t$ algorithm~\cite{Ellis:1993tq}
\item Anti-$k_t$ algorithm ~\cite{Cacciari:2008gp}
\item Cambridge/Aachen Algorithm (CA)~\cite{Dokshitzer:1997in}
\end{itemize}
These algorithm calculates in the following way. At first, the inter particle distance 
and the beam distance are measured via,
\begin{equation}
d_{ij}=\text{min}({p^{2p}_{T_i},p^{2p}_{T_j}}), ~~~~\text {for all} (i,j) 
\end{equation}
\begin{equation}
d_i=p^{2p}_{T_i} R^2 ~~~~~\text {for all} i.
\end{equation}
Then among $d_{ij}$ and $d_i$, whichever is smallest is chosen. If $d_{ij} < d_i$, 
the $i$-th and $j$-th jets are combined in to a new jet $k$. If $d_{ij} > d_i$, 
then object $i$ is considered as a jet.
Here $p$ is an arbitrary parameter, which is different in different algorithm. 
In $k_t$ algorithm, $p=1$, CA algorithm $p=0$, Anti-$kt$ algorithm $p=-1$.
Among these algorithms, $k_t$ algorithm shows the best sensitivity towards soft emissions.

Cone algorithm~\cite{Blazey:2000qt} is based on the idea of finding a perfect cone. At a given centre 
in the ($y-\phi$) plane, the 4 momentum of all particles are summed at a given jet radius ($R$)
and the cone is stable if the sum of all particle four momentum coincides with the direction 
of the cone point. Examples are: {\it Midpoint} type and {\it Jetclue} algorithms.
At the colliders, recombination algorithms are generally preferred 
because these algorithms are faster than the cone algorithm.
\section{ Jet Substructure}
The statistical nature of the jet reconstruction algorithm forbids it from exact differentiation 
among the jets. 
The QCD jets are of two types, quark jets or gluon jets. 
The probability for a gluon to radiate a gluon is larger 
by a factor of $C_A /C_F = 9/4$ (ratio of QCD color factors) than a quark having the 
same energy fraction and angle. Hence, gluon jets
tend to have more constituents and a broader radiation
pattern than quark jets. For a detailed discussion on 
quark-gluon discrimination follow~\cite{Gallicchio:2011xq}.

Moreover, the final state particles will have a larger $p_T$ if the center of mass energy 
increases (For example: LHC at 14 TeV, HE LHC upgrade). Then, these particles will behave like 
{\it boosted jet}.
It becomes very hard to distinguish or isolate the boosted $W$, $Z$, $H$ from the quark and gluon jets, 
which are considered as {\it standard jets}. There are many techniques to 
distinguish between different types of jets.
The job of different jet substructure processes are to identify the type of the jet first, 
then determine its {\it pronginess}, that is how many hard cores are there in a jet.  
For example, QCD jets are one prong, $W$/$Z$/$H$ are two prong, and top is a three 
prong jet. The bosonic jets, $W$/$Z$/$H$, are called {\it fatjets} because they are large radius jets, 
where the decay products of them are all inside the jet cone.
Another difference between them 
is that the QCD jets will carry more soft gluon radiation than the $W$/$Z$/$H$ jets. In order to 
cut the effect of soft backgrounds of these fatjets {\it groomers} are used. Let us look at 
some of the popular techniques. 
\subsection{Techniques}
\paragraph{Pruning and Trimming}

Pruning is one of the jet-grooming methods~\cite{Ellis:2009su} to remove
the constituents from the jets that carry no significant or useful
information.
Mathematically, at each merging
step ($i + j \rightarrow k$), we define two constraints given by:
\begin{itemize}
\item Softness: $p_j/p_k < z_{\rm cut}$ for $(p_j < p_i)$ 
\item Separation: $\Delta R_{ij} > R_{\rm cut} $
\end{itemize}
If both these
conditions are met, then we prune (remove) the constituent $k$ and
proceed for next merging. A larger (smaller) value for $z_{\rm cut}
\, (R_{\rm cut})$ will result in more aggressive pruning. The
level of pruning is determined by the less aggressive of these two
parameters. 
The pruned jet mass, $m_j$, is computed from
the sum of the four-momenta of the components that remained after
pruning. For example, in order to distinguish a fat $Z$ from the 
QCD background jets jetmass is a very useful parameter. 
The QCD jets will have a lower jetmass but for a fat $Z$,
the jetmass will coincide with $m_Z$.
On the other hand, {\it trimming} or {\it filtering} use a top down approach,
where it reclusters the constituents of a jet with jet algorithm with a small 
radius ($R_{trim}$), and keeps only the jets with larger $p_T$ or 
larger than a fraction $f_{\text{trim}}$.

\paragraph {Mass drop tagger and SoftDrop}
In this process, after reclustering the jet one looks at the last step ($i+j\rightarrow k$) of the iteration 
and checks,
\begin{itemize}
\item Whether the reconstructed mass is less than the initial masses (hence the 
name mass drop) or not i.e, ${\rm max}(m_i,m_j)<\mu_{\rm cut} ~ m_{k}$
\item The splitting is symmetric or not, i.e, 
${\rm min}(p_{T_i}^2,p_{T_j}^2)\Delta R_{ij}^2>y_{\rm cut} ~ m_{k}^2$.
\end{itemize}
When both criteria are met, we keep ``$k$'', if not, the heavier jet between $i$ and $j$ is
selected and the procedure is repeated. 
The last condition modifies to 
${\rm min}(p_{T,i},p_{T,j})>z_{\rm cut} p_{T_k}$ in {\it modified mass drop Tagger (mMDT)}
which leads to same analytical behaviour.
In {\it SoftDrop}, the symmetry condition is replaced by a more general form\\
\begin{equation}
\frac{{\rm min}(p_{T_i},p_{T_j})}{p_{T_i}+p_{T_j}} > z_{\rm cut} \Big(\frac{\Delta R_{ij}}{R}\Big)^\beta
\end{equation}
Here the parameter $\beta$ controls the the aggressiveness of the grooming.
In general, $\beta=-1$ is used. A more negative value will lead to a more aggressive grooming.
A good example for these methods is CMS top tagger~\cite{CMS:2009lxa}, HEP Top Tagger~\cite{Plehn:2010st}.

\subsection{Observables}
In order to differentiate the jets originating from the signal from 
SM background, different observables are used. I discuss only a subset of them.
For a more robust discussion one can follow 
Ref:~\cite{Marzani:2019hun,Abdesselam:2010pt,Altheimer:2012mn,Adams:2015hiv}.

\paragraph{Jetmass}
Jetmass distributions are generally calculated via either the grooming method
or via trimming or pruning method and by doing resummation at all orders.
QCD jets acquire mass from emissions during parton shower. 
If we consider the case of $q\rightarrow q g$ for small $R$, 
the average jetmass squared mass can be written as~\cite{Shelton:2013an},
\begin{equation}
<m^2> \approx \dfrac{\alpha_s}{\pi} \dfrac{3}{8} C_F p_T^2 R^2
\end{equation}
So, the jet mass scales as $p_T$ and increases linearly with $R$. 
If we do a similar analysis for the splitting, $g \rightarrow  q q$, we get:
\begin{equation}
<m^2> \approx \dfrac{\alpha_s}{\pi} \dfrac{1}{20} C_A \ p_T^2 R^2
\end{equation}
If we now look at the masses in case of fatjets like Z, Higgs etc., 
with the splitting given by $k \rightarrow i j $, we can write the jet mass as:
\begin{equation}
m^2 \approx 2p_i . p_j \approx p_{T,i} p_{T,j} \Delta R_{ij}^2 = z(1-z) p_{T,k}^2 \Delta R_{ij}^2
\end{equation}  
$z=Eq/Ep$,  $\Delta R_{ij} = 2m/p_T$ fraction of parton energy.
These relation show that the jet mass has a dependence on the jet radius,  
which is discussed explicitly in~\cite{Choudhury:2021nib}.
\paragraph{N-subjettiness}
$W$, $Z$ or a Higgs
boson fatjets are well studied in ~\cite{Asquith:2018igt,Sirunyan:2017usq,Sirunyan:2016ipo}. 
If the fatjet or a jet is resolved into subjets then a good measure of the number of subjets is given by
N-subjettiness~\cite{Thaler:2010tr} defined as,
\beq
\tau_N=\dfrac{1}{d_0} \sum_{k} p_{\rm T,k} \, {\rm min}\Big(\Delta R_{1k},\Delta R_{2k},....\Delta R_{Nk}\Big)
\eeq
where $N$ is the number of subjets of the jet to be
reconstructed. $k$ runs over constituent particles in a
given jet, and $p_{\rm T,k}$ are their transverse momenta and $\Delta
R_{j,k}$ the angular separation between a candidate subjet $j$ and a constituent particle $k$. Furthermore,
\beq
d_0=\sum_{k} p_{\rm T,k} R_0
\eeq
where $R_0$ is the characteristic jet-radius. 
Physically, $\tau_N$ provides a dimensionless measure of whether a jet
can be regarded to be composed of $N$-subjets. 
If the value of $\tau_N$ is small, then there are $N$ or less than 
$N$ subjets are involved. Boosted particle such as a Higgs or $W/Z$ show two prong nature 
as they decay and will have large $\tau_1$ and small $\tau_2$. 
QCD jets which have small $\tau_2$ will typically have smaller $\tau_1$,
due to their huge energy spread. QCD jets which have large $\tau_1$ are 
considered as diffused jets and will have larger $\tau_2$ as well compared to the signal.
For a detailed understanding check ~\cite{Thaler:2010tr}.

\begin{figure}
\begin{center}
\includegraphics[width=6.5cm,height=6.3cm]{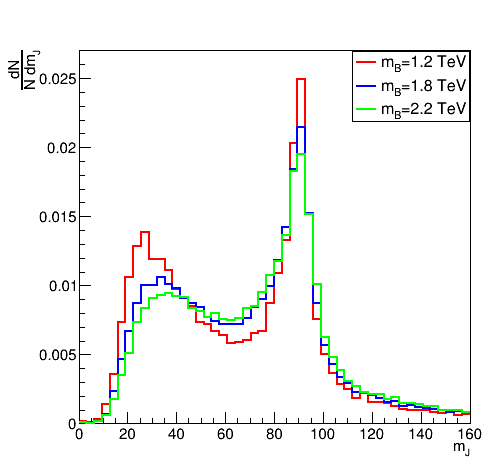}
\includegraphics[width=6.5cm,height=6.3cm]{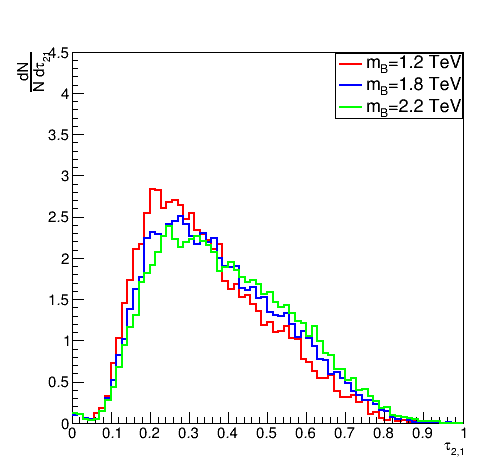}
\end{center}
{\caption{\label{fig:1}} (left) Mass distribution($m_j$) for $Z$-fatjet. (right)$\tau_{21}$ distribution of the $Z$-fatjet. Plots are taken from the study in Ref:~\cite{Choudhury:2021nib}.} 
\end{figure}

In particular, ratios of $N$ and $(N-1)$th subjettiness, are powerful discriminants between jets predicted
to have larger or fewer number of internal energy clusters. The ratio is given by,
\beq
 \tau_{N~N-1}=\tau_N/\tau_{N-1}
\eeq
Clearly, $\tau_{21}$ will be a good discriminator if we have a signal rich with fatjet $Z$ and 
dominating QCD backgrounds. Jets coming from the hadronic
decays of the $Z$ tends to have lower values for the ratio $\tau_{21}
\equiv \tau_2/\tau_1$ and, hence, this is a good discriminator. In Fig:\ref{fig:1}, I have shown 
the distribution of the jetmass, $\tau_{21}$ and the correlation between 
the jet mass and $\tau_{21}$ for this specific example of $Z$-fatjet, emerging from 
a vectorlike Bottom quark ($B$).
Similarly, for a signal enriched with boosted top, $\tau_{31}$ will be a better 
discriminator, which is implemented in HEP Top Tagger~\cite{Plehn:2010st}.
\paragraph{Energy correlation variables: $C_2^{\beta}$,  $D_2^{\beta}$,  $N_2^{\beta}$}\
This a a more advanced variable in a sense that unlike N-subjettiness, there is no 
different minimisation procedure in the algorithm, in order to get the direction of the 
subjets. More over, energy correlation variables~\cite{Larkoski:2013eya} are insensitive to the recoil due 
to soft emissions. Soft wide angle radiation displaces the hard jet core from
the jet axis to balance the angular momentum. N-subjettiness variable is 
sensitive to this displacement since they are measured with respect to the jet center. 

As the $(N+1)$ point correlation functions are sensitive when the jet substructure is $N$-prong, 
the energy correlation variable is defined as $C_{N}^{\beta}$. As we saw in the ratio of subjettiness, 
we use $C_{1}^{\beta}$ for QCD jets, $C_{2}^{\beta}$ for bosonic jets, $C_{3}^{\beta}$ for top quark.
Here $\beta$ is the angular exponent.
The corresponding energy co relation variables are,\\
\begin{equation}
E^{\beta}_{2}= \sum_{i<j}p_{T_i} p_{T_j} R_{ij}^{\beta} ~~~ E^{\beta}_{3}= \sum_{i<j<k}p_{T_i} p_{T_j} p_{T_k} (R_{ij}R_{ik}R_{jk})^{\beta}
\end{equation}
\begin{equation}
E^{\beta}_{z}= \sum_{i<j<....<z}p_{T_i} p_{T_j}... p_{T_z} (R_{ij}R_{ik}....\text{all possible combinations})^{\beta}
\end{equation}
Just as we saw in case of N-subjettiness ratio, here also ratio of these variables are 
important. These are defined as, 
\beq
r_N^{\beta}= \frac{E(N+1)}{E(N)}~~~~~~~C_N^{\beta}=\frac{r_N^{\beta}}{r_{N-1}^{\beta}}
\eeq
$C_N$ effectively measures higher-order radiation
from leading order (LO) substructure. For a system with $N$ subjets, the LO substructure
consists of N hard prongs, so if $C_N$ is small, then the higher-order radiation must be soft or
collinear with respect to the LO structure. If $C_N$ is large, then the higher-order radiation
is not strongly-ordered with respect to the LO structure, so the system has more than $N$
subjets. Thus, if $C_N$ is small and $C_{N}^{1}$ is large, then we can say that a system has $N$
subjets. 
\section{Conclusion}
In this review, I have tried to briefly summarise the main lessons of jet substructure. 
In the past few years, there have been many theoretical developments in this subject. 
Moreover, it has seen a huge success at LHC experiments to help to increase the 
sensitivity of new physics. As the energy of LHC is planned to 
increased in future, the jet substructure methods will be of huge 
importance in recent future to study the boosted objects specifically. 
Advanced techniques such as Neural Network and Machine Learning
are opening new doors to study the new physics by making jet substructure more robust. 
Moreover, there will be implementation of jet substructure in future collider experiments as well.
Hence, I believe this era is the golden era for jet substructure in High Energy Physics and this short review 
might help the beginners to get a basic idea on jet substructure.
\bibliographystyle{JHEP}
\bibliography{vecB.bib}
\end{document}